\documentstyle[11pt,IAU207_pasp,twoside]{article}
\def\edcomment#1{\iffalse\marginpar{\raggedright\sl#1\/}\else\relax\fi}
\reversemarginpar

\begin{document}
\title{Second-parameter Globular Clusters in the Milky Way\\
and in M33 as Tracers of Mass Loss from M31 in the\\
Early Epoch?}
\author{Valery V. Kravtsov}
\affil{Sternberg Astron. Ins., 13, University Ave., Moscow 119899,
Russia}

\begin{abstract}
I suggest that the bulk of the "young", second-parameter globular clusters
(SPGCs) observed in the outer halo of our Galaxy and recently found in other
massive spiral of the Local Group (LG), M33, may have originated due to mass
outflow from M31 and subsequent accretion of gas on the Galaxy and M33 in
the early epoch.
\end{abstract}

\section{Problem and Main Questions}

There are contradictory conclusions about the dwarf spheroidal (dSph)
galaxies' tidal dissolution and its rate, as well as a number of observational
constraints on this process in the Galaxy (e.g., van den Bergh 1996; Gilmore
et al. 1999; Shetrone et al. 2000). Moreover, preliminary results on the CMD
study of GCs in massive spirals of LG suggest a fraction of SPGCs to be
at least low  among GCs in M31 (Rich et al. 2001) and to be high as for M33
GC population (Sarajedini et al. 1998). Hence there appear doubts about
tidal disruption/stripping of dSphs as universal mechanism of and main
contributor to a formation of populations of the SPGCs in the above three
galaxies. Note also that Chandar et al. (2001) have estimated the total
number of GCs in M 33 to be $75 \pm14$ which gives a specific frequency
significantly higher than in other late-type spirals.

By now, it is established that the Galactic population of SPGCs shows a
number of basic distinctive characteristics as compared with
those of other GCs in the halo of the Milky Way (e.g., Da Costa 1994; van
den Bergh 1996). Some of these characteristics imply that the
Galactic SPGCs are very likely to have an accretive nature.
However, if a tidal dissolution of dSphs is not main and the only
contributor to the formation of population(s) of the SPGCs
in our Galaxy (and in M33) then the following vexed questions arise.

{\it What is a material accreted and where did
it originate and come from?
Why has the formation of SPGCs lasted after
formation of the bulk of halo objects in rapid collapse?}

\section{Putative Role of the Mass Outflows}

The high-redshift Layman break galaxies (LBGs) are suggested to be
progenitors of the present-day massive spheroids being in particular
components of luminous early-type spirals (Friaca \& Terlevich 1999).
As summarized by Heckman (2000) starburst-driven galactic superwinds
observed in LBGs play important role in the mass outflow from them. During
main, most powerful episode ({\bf $\bf{\sim 2}$ Gyr}) of formation of
the Pop. II stars in a star forming galaxy like M31, its outflows may carry
off the galaxy the total mass of gas as large as mass of the stellar bulge
and spheroid, with typical speeds ranging from {\bf a few} ${\bf 10^{2}}$
to {\bf ${\bf10^{3}}$ km/s} (Pettini et al. 2001). Some a few tens percent
of this mass (${\bf \Delta M \sim 10^{11} M_{{\bf  \odot}}}$ or so) with
velocities exceeding the escape velocity of M31 could have quited the
galaxy, portion of which might be accreted on our Galaxy (and M33). If we
accept present distance between the Galaxy and M31
({\bf ${\bf \sim0.7}$ Mpc}), ratio of their mass
(${\bf {{\bf M_G}}/{{\bf M_A}} \sim1/1.5}$), and also a mean velocity of
expanding gas to be of order {\bf ${\bf 400}$ km/s}, then we easily find
that a time spent by gas to reach a region of the Lagrangian point between
the two galaxies, may be as long as {\bf ${\bf \sim1}$ Gyr}.
{\bf An upper limit of mass ${\bf \delta M}$ of gas accreted on the Galaxy}
is estimated from a formula obtained (in the approach of gas accretion in
gravity potential $\varphi \sim R^{-1}$ of a field of less massive companion
in a binary system) and kindly offered by Postnov
(2000, private communication):

$$\delta M \sim{{1}\over{4}}{ \bigg({{M_G} \over{M_A}} \bigg)}^2\Delta M
\sim{{1}\over{10}}\Delta M \sim10^{10} M_{ \odot}$$

Even if a mass of really accreted gas was an order of magnitude lower than
the upper limit estimated and if in turn approximately ten percent of
this mass was converted into objects, i.e., {\bf ${\bf \sim10^{8} M_{\odot}}$}, that
would be equal to a {\bf total mass} of the bulk of {\bf SPGCs}, of a few
{\bf dSphs}, and of some portion of {\bf field stars} which may have formed
due to such a process in the outer halo of our Galaxy.

Similarly, it is possible to conclude that objects assumed to form from a gas
accreted onto M33 might comprise a total mass of order
{\bf a few ${\bf 10^{6} M_{\odot}}$}.
Some {\bf 10 to 20 GCs} (with fraction of {\bf field stars} formed) with
cluster's typical mass of order $10^{5} M_{\odot}$ may amount just to this
mass.

\acknowledgements

I am deeply thankful for the IAU travel grant. Also, I am very
grateful to Dr. A.A. Suchkov for his encouraging and valuable comments on the
scenario presented, and to Drs. K. Postnov, Yu.A. Shchekinov, Yu.N. Efremov,
N.I. Shakura, and M.V. Sazhin for helpful discussions.

\end{document}